\documentclass[useAMS,usenatbib,twocolumn]{mnras}
\usepackage[figuresright]{rotating}
\usepackage{lscape}

\usepackage{graphicx}
\usepackage{amsmath}
\usepackage{amssymb}
\usepackage{multirow}
\usepackage{setspace}
\usepackage{bigdelim}
\usepackage{bigstrut}
\usepackage[rflt]{floatflt}	
\usepackage{wrapfig}
\usepackage{caption}
\usepackage{multicol}

\usepackage{indentfirst}

\title[Indirect Influence of Quasars on Reionization]{The Indirect Influence of Quasars on Reionization}
\author[Seiler et al]{Jacob Seiler$^{1,2}$\thanks{E-mail:
		jseiler@swin.edu.au}, Anne Hutter$^{1,2}$, Manodeep Sinha$^{1,2}$, Darren Croton$^{1,2}$ \\
\\
$^{1}$Centre for Astrophysics \& Supercomputing, Swinburne University of Technology, PO Box 218, Hawthorn, Victoria 3122, Australia \\
$^{2}$ARC Centre of Excellence for All Sky Astrophysics in 3 Dimensions (ASTRO 3D)}

\begin{document}


\pagerange{\pageref{firstpage}--\pageref{lastpage}} \pubyear{2002}

\maketitle

\label{firstpage}

\begin{abstract}

The exact role of quasars during the Epoch of Reionization remains uncertain. With consensus leaning towards quasars producing a negligible amount of ionizing photons, we pose an alternate question: Can quasars \textit{indirectly} contribute to reionization by allowing ionizing photons from stars to escape more easily?  Using the Semi-Analytic Galaxy Evolution model to evolve a galaxy population through cosmic time, we construct an idealized scenario in which the escape fraction of stellar ionizing photons ($f_\mathrm{esc}$) is boosted following quasar wind events, potentially for several dynamical times.  We find that under this scenario, the mean value of $f_\mathrm{esc}$ as a function of galaxy stellar mass peaks for intermediate mass galaxies.  This mass dependence will have consequences for the $21$cm power spectrum, enhancing power at small scales and suppressing it at large scales.  This hints that whilst quasars may not directly contribute to the ionizing photon budget, they could influence reionization indirectly by altering the topology of ionized regions.

\end{abstract}


\section{Introduction}

The Epoch of Reionization represents a transition between the neutral, post-recombination Universe and the highly ionized one that we observe today.  As the first stars form, they release photons which gradually ionize the neutral hydrogen within the intergalactic medium (IGM) by $z \sim 6$ \citep{Fan,Becker2015}.  The time and spatial evolution of ionized regions during reionization depends on the fraction of ionizing photons that escape their host galaxies into the IGM.  The functional form and value of this parameter, the escape fraction $f_\mathrm{esc}$, remains highly uncertain since direct observations are impeded by the partially neutral IGM.

One of the most common implementations for the form of $f_\mathrm{esc}$ is to assume a constant value for all galaxies over cosmic time \citep[e.g.,][]{Iliev2007, Raicevic2011, Hutter2014} with some authors investigating the impact of a redshift-dependent $f_\mathrm{esc}$ \citep[e.g.,][]{Kuhlen2012, Kim2013a}.  By accounting for the properties and physical processes of galaxies and their host halos, it has also been postulated that $f_\mathrm{esc}$ may scale either positively \citep{Gnedin2008a,Wise2009} or negatively \citep{Yajima2011,Ferrara2013,Kimm2014,Xu2016} with halo mass.   Using zoom-in radiation-hydrodynamic simulations, \cite{Kimm2017} find that the value of $f_\mathrm{esc}$ varies over time and is highly sensitive to the physical processes that disperse dense gas pockets within the galaxy.  Work by \cite{Paardekooper2015} highlight the importance of the covering fraction of galaxy gas, finding that ionizing radiation escapes through one or two patches on the sky; in all other directions $f_\mathrm{esc}$ is approximately zero.  These results indicate that the escape fraction depends complexly upon the distribution and density of gas within each individual galaxy.

Another open question focuses on the sources of ionizing photons during the Epoch of Reionization.  Whilst general consensus leans towards star-forming galaxies dominating the ionizing photon budget \citep[e.g.,][]{Robertson2015,Finkelstein2015}, the detection of faint active galactic nuclei at $z>4$ by \cite{Giallongo2015} has reignited discussion regarding the contribution of quasars to reionization.  Although after accounting for these extra faint objects and matching their models to measurements of the optical depth by \cite{Planck2016b}, many authors have concluded that quasars contribute a negligible number of ionizing photons compared to star-forming galaxies \citep[e.g.,][]{Khaire2015, Hassan2016,Qin2017,Parsa2018}.\footnote{However also see work claiming that star-forming galaxies alone do not produce sufficient ionizing photons to complete reionization \citep[e.g.,][]{Kollmeirer2014, Grazian2017, Madau2017}.} 

In this letter, we assess the impact of quasars on the reionization topology through their ability to allow the easy escape of stellar ionizing photons from galaxies for a number of dynamical times.  For our work, we use the ionizing photon distribution as a function of galaxy stellar mass as a proxy for reionization topology.   Our fundamental question differs from other works as we consider the impact of quasars on the surrounding galaxy gas rather than focusing on their intrinsic emission of ionizing photons. 

This paper is organised as follows: In Section \ref{sec:Modelling} we describe our underlying N-body simulation and semi-analytic model which evolves a galaxy population from $z=15$-$6$.  In Section \ref{sec:fesc}, we construct a toy model that links the value of $f_\mathrm{esc}$ to quasar winds to allow ionizing photons to easily escape. We then analyze the evolution of the ionizing emissivity and form of $f_\mathrm{esc}$ as a function of galaxy stellar mass under this quasar activity scenario.  We conclude this Section with a comment on the impact our toy model has on the $21$cm power spectrum.  We conclude in Section \ref{sec:Conclusion}.

Throughout this letter we adopt the cosmological values $\left(h, \Omega_\mathrm{m}, \Omega_\Lambda, \sigma_8, n_s\right)= \left(0.681, 0.302, 0.698, 0.828, 0.96\right)$ consistent with \cite{Planck2016b} and use a \cite{Chabrier2003} initial mass function (IMF).

\section{Simulation and Semi-Analytic Modelling}  \label{sec:Modelling}

\subsection{N-Body Simulation}

For our work, we use the collisionless N-Body simulation \textit{Kali}. To ensure our simulation represents a mean (i.e., not over-dense or under-dense) region of the Universe, we generate $1000$ initial conditions, each corresponding to a different random number seed, and choose the initial condition that has the smallest root-mean-square (rms) from the theoretical linear power spectrum.  The initial conditions are generated using second order Lagrangian perturbation theory with the code \texttt{2LPTic} \citep{Scoccimarro1998, Crocce2006}.  

\textit{Kali} contains $2400^3$ dark matter particles within a $160$ Mpc side box and is evolved through time using \texttt{GADGET-3} \citep{Springel2005}.  This box size is large enough to contain the largest ionized regions of $10$-$20$ Mpc \citep{Geil2016} whilst the particle mass resolution of $1.15 \times 10^{7} \: \mathrm{M}_\odot$ can sufficiently resolve halos of mass $\sim 4 \times 10^8\:\mathrm{M}_\odot$ which are thought to drive reionization \citep{Paardekooper2013}.  The dark matter particles were evolved with $98$ snapshots of data stored between redshifts $z$ = $30$-$5.5$ (equally spaced every $10$Myr). 

We identify halos in the dark matter distribution using \texttt{SUBFIND} \citep{Springel2001} with a friends-of-friends linking length chosen to be $0.2$ times the mean inter-particle separation.  Halo merger trees are generated with \texttt{GBPTREES} \citep{Poole2017}.

\subsection{Semi-Analytic Modeling} \label{sec:SAM}

In our work, we use the Semi-Analytic Galaxy Evolution (\texttt{SAGE}) model to evolve a galaxy population across cosmic time \cite[][hereafter C16]{Croton2016}.  This model includes baryonic accretion, cooling, star formation, gas ejection due to supernova feedback, AGN feedback through `radio mode' heating and `quasar mode' gas ejection, and galaxy mergers. 

With respect to C16, the only model prescription that has been changed is the supernova feedback scheme.  This revised scheme is more appropriate at high redshift where the snapshot timescales are smaller than the lifetime of supernova candidate stars.  Unlike C16, where after each star formation episode a fixed fraction of stars instantly explode, we closely follow \cite{Mutch2016} and release energy from supernova explosions gradually over a number of subsequent snapshots.  

As our ultimate goal is to create a toy model in which the escape fraction of ionizing photons is enhanced by quasar activity, we briefly describe the implementation of gas ejection due to quasar winds within the \texttt{SAGE} model.  Following a merger event,  rapid accretion of cold gas onto the central galaxy's black hole is triggered with a rate dependent upon the mass ratio of the merging galaxies. C16 adopts a simple phenomenological model whereby following this accretion, a quasar wind sweeps across the galaxy with energy proportional to the accreted mass. If the energy of this quasar wind is greater than the thermal energy in the cold disk, the cold gas and associated metals are blown out of the galaxy.  If the quasar energy is greater than the total thermal energy of both the cold gas and hot halo gas, the quasar wind ejects all gas and metals from both cold and hot gas reservoirs.  This total ejection is consistent with work showing mass outflows on the order of $10^3\mathrm{M}_\odot\mathrm{yr}^{-1}$ which, over our $10^7$yr simulation timestep, is sufficient to eject all gas within the galaxy \citep{Gabor2014,Bieri2017}.

\begin{figure*}
	\includegraphics[scale = 0.4]{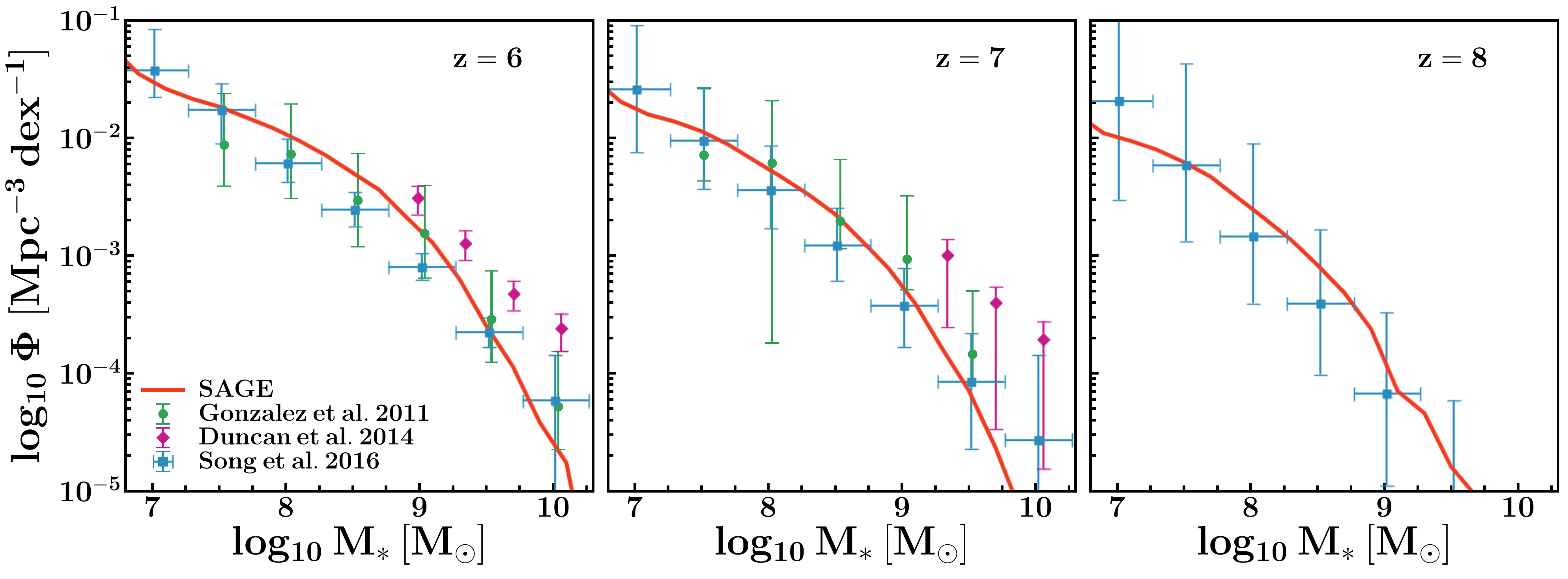}
	\caption{High redshift ($z = 6$-$8$) stellar mass function produced by our modified version of \texttt{SAGE} that includes delayed supernova feedback.  We tune the input parameters to match the observational data points from \citet{Gonzalez2011}, \citet{Duncan2014} and \citet{Song2016}.  All observations have been corrected to a Chabrier IMF and $h = 0.698$.}
	\label{fig:StellarMass}
\end{figure*}

We calibrate the \texttt{SAGE} parameters manually to match the high redshift stellar mass function using \cite{Gonzalez2011}, \cite{Duncan2014} and \cite{Song2016} between $z=6$-$8$, shown in Figure \ref{fig:StellarMass}. This involved altering the following C16 parameters: the star formation rate $\alpha_\mathrm{SF}$ from $0.05$ to $0.01$, the recycle fraction from $0.43$ to $0.25$ and the quasar mode ejection coupling $\kappa_\mathrm{Q}$ from $0.005$ to $0.02$.  We use the \cite{Mutch2016} mass loading and energy coupling constants for supernova feedback.	From Figure \ref{fig:StellarMass} we see that the stellar mass function fits the observations well over all redshifts, highlighting the robustness of the \texttt{SAGE} model during the Epoch of Reionization.

\section{Escape Fraction As A Function of Quasar Activity} \label{sec:fesc}

In order to explore the impact of quasars on the reionization topology, we compare the results of a scenario in which quasar winds allow the easy escape of stellar ionizing photons to a constant value of $f_\mathrm{esc}$.  In this Section we first describe our $f_\mathrm{esc}$ prescription that depends upon quasar activity within the \texttt{SAGE} model, to which we refer to as the ``quasar activity scenario'' below.   Secondly, we elaborate on our numeric $f_\mathrm{esc}$ parameters chosen to match the observed ionizing emissivity evolution.  Finally, we discuss the variation of mean $f_\mathrm{esc}$ values across galaxy stellar mass for the quasar activity scenario and its impact on the observed $21$cm power spectrum compared to the constant $f_\mathrm{esc}$ scenario.

\subsection{The Toy Model}

Motivated by findings that the escape fraction of ionizing photons is dictated by local galaxy processes altering the gas distribution \citep[e.g.,][]{Paardekooper2015, Kimm2017}, we employ a toy model which links a galaxy's value of $f_\mathrm{esc}$ to its quasar activity. In the \texttt{SAGE} framework this is achieved by boosting the value of $f_\mathrm{esc}$ from $0.15$ to $1.0$ for a short period following a quasar wind event that ejects all cold and hot gas within the galaxy as the lack of impeding gas allows easy photon escape.  This period of free photon streaming will last until a significant amount of gas returns to the galaxy, through either reincorporation or baryonic infall.  We parametrize this timescale using the dynamical time of the host halo.  After this time, $f_\mathrm{esc}$ drops back to its baseline value of $0.15$. For our toy model we allow the boosted value of $f_\mathrm{esc}$ to last for two and a half dynamical times.  We motivate our choice of baseline $f_\mathrm{esc}$ value and number of dynamical times in Section \ref{sec:Emissivity}. 

\subsection{Ionizing Emissivity \label{sec:Emissivity}}

The number of ionizing photons that contribute to the ionization of the IGM is determined by the number of ionizing photons intrinsically produced by each galaxy ($N_{\gamma,i}$) and their value of $f_\mathrm{esc,i}$, 

\begin{equation}
N_\mathrm{ion} = \sum_i f_{\mathrm{esc},i} N_{\gamma,i}.
\end{equation}

In this work, we link $N_{\gamma,i}$ to a galaxy's star formation rate and metallicity.  This is achieved by computing spectra for multiple star formation rate and metallicity bins using the stellar population synthesis code \texttt{STARBURST99} \citep{Leitherer1999} and fitting a linear equation for $N_{\gamma,i}$ as a function of galaxy star formation rate.

For the quasar activity scenario, we calibrate the aforementioned baseline value of $f_\mathrm{esc}$ and boost timescale to match the ionizing emissivity found in \cite{Bouwens2015}, using values of $0.15$ and $2.50$ dynamical times respectively. For comparison, we include a reference scenario that uses a constant value of $f_\mathrm{esc} = 0.25$ at all times, chosen again to match the ionizing emissivity of \cite{Bouwens2015}.  Using the semi-numerical code \texttt{cifog} \citep{Hutter2018} to simulate the reionization of the IGM, these calibrations yield optical depth values of $\tau = 0.058$ for both reference and quasar activity scenarios, which agree with the \cite{Planck2016} measurements of $\tau = 0.058 \pm 0.012$. 

\begin{figure}
	\hspace*{-0.3in}
	\includegraphics[scale = 0.38]{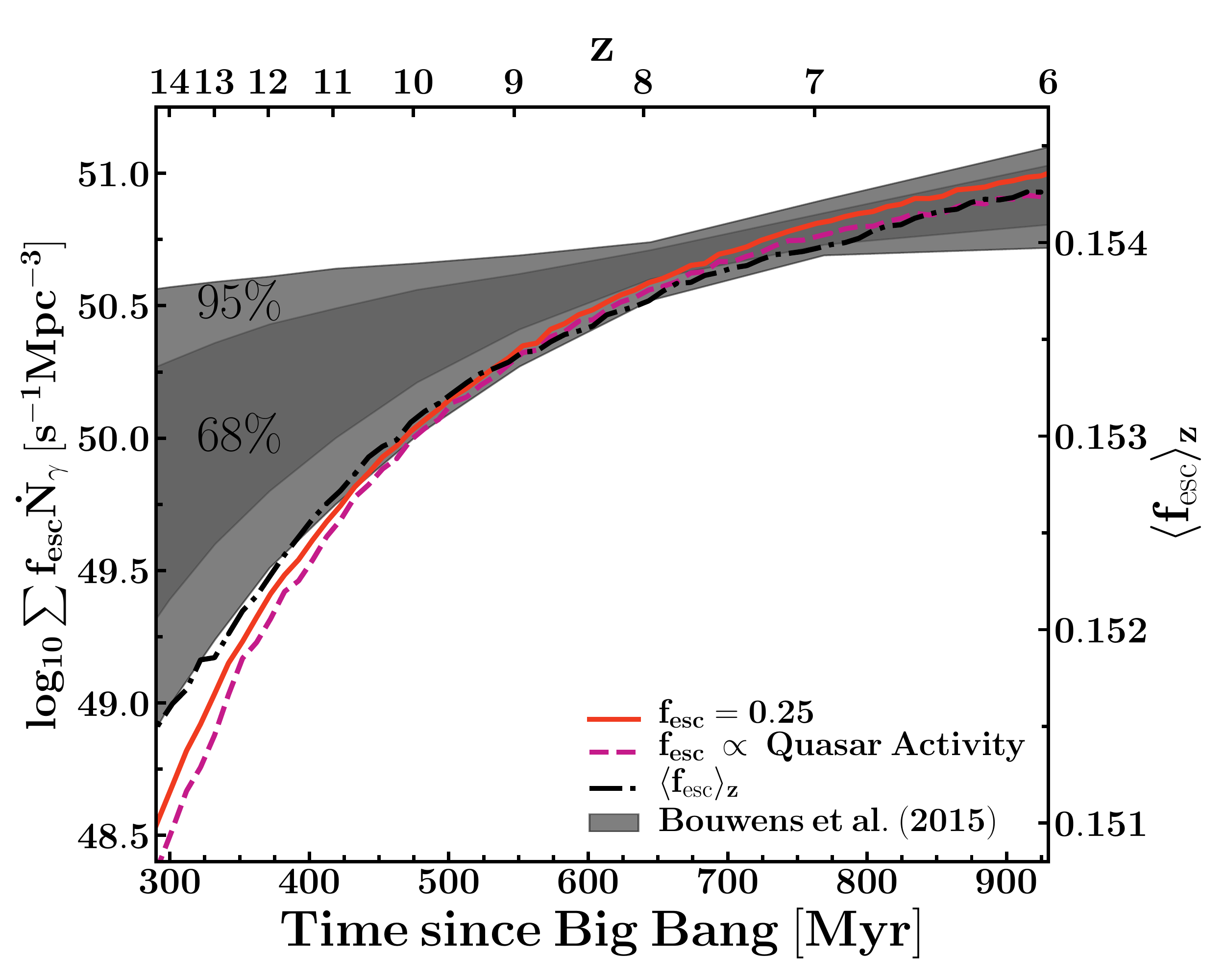}
	\vspace*{-0mm} 
	\caption{Left  axis: Ionizing emissivity for the quasar activity scenario and a constant $f_\mathrm{esc} = 0.25$ scenario for reference.  The shaded contours show the derived $68\%$ and $95\%$ confidence intervals for the ionizing emissivity inferred using the Thomson optical depth, quasar absorption spectra and the prevalence of Ly$\alpha$ emission in $z = 7$-$8$ galaxies \citep[][Table 2]{Bouwens2015}. Right axis: The average $f_\mathrm{esc}$ value as a function of redshift for the quasar activity scenario.	For this cenario each galaxy has a baseline $f_\mathrm{esc}$ value of $0.15$ with a boosted $f_\mathrm{esc}$ value of $1.00$ for two and a half dynamical times following a quasar wind ejecting all gas from a galaxy.}
	\label{fig:Nion}
\end{figure}

Figure \ref{fig:Nion} compares the evolution of the ionizing emissivity for our two $f_\mathrm{esc}$ scenarios with the inferred estimates of \cite{Bouwens2015}.  We see that both models are consistent with these estimates over majority of the redshift range of reionization.  However, at very early times ($z \approx 14$), we see that both models produce too few ionizing photons. While this could be accounted for by choosing larger values of $f_\mathrm{esc}$ (e.g., a constant value of $f_\mathrm{esc} = 0.30$ and a baseline $f_\mathrm{esc} = 0.20$ for the quasar activity scenario), such a choice would result in too many ionizing photons being produced at later times and deviating from observational estimates.  At redshift $z = 14$, low mass objects emit the majority of the ionizing photons, suggesting that either the \texttt{SAGE} model is forming too few low mass galaxies or that the value of $f_\mathrm{esc}$ should be larger at early times.  We are disinclined to believe the former explanation as the \texttt{SAGE} stellar mass function (Figure \ref{fig:StellarMass}) shows that we form a sufficient number of low mass galaxies.  The latter explanation aligns with work showing that $f_\mathrm{esc}$ should increase for low mass halos \citep[e.g.,][]{Yajima2011,Ferrara2013,Kimm2014,Xu2016}, hinting that an extra physical mechanism, such as linking the value of $f_\mathrm{esc}$ to supernova activity, is required to boost the value $f_\mathrm{esc}$ within low mass galaxies.  Furthermore, boosting $f_\mathrm{esc}$ for a subset of galaxies can have a significant impact on the ionizing emissivity, as shown by the right axis of Figure \ref{fig:Nion}.  We see that despite the average value of $f_\mathrm{esc}$ being well below the constant $f_\mathrm{esc} = 0.25$ reference scenario, we obtain a comparable number of escaping ionizing photons.

Finally, the mass resolution of \textit{Kali}, resolving halos of $\sim 4 \times 10^8\:\mathrm{M}_\odot$, neglects the contribution of galaxies living within extremely low mass halos. As shown by \cite{Paardekooper2013} and \cite{Kimm2017}, at early times ($z>10-11$) a significant portion of ionizing photons ($\sim 20-40\%$) are potentially emitted by these galaxies and may explain our discrepancy with the \cite{Bouwens2015} results at high redshift.  Furthermore, our model does not account for  pop-III star formation which could also contribute a considerable number of ionizing photons to reionization \cite[e.g.,][]{Chen2017,Kimm2017}.

\subsection{Mean $f_\mathrm{esc}$ as a Function of Stellar Mass}

\begin{figure}
	\hspace*{-0.0in}
	\includegraphics[scale = 0.4]{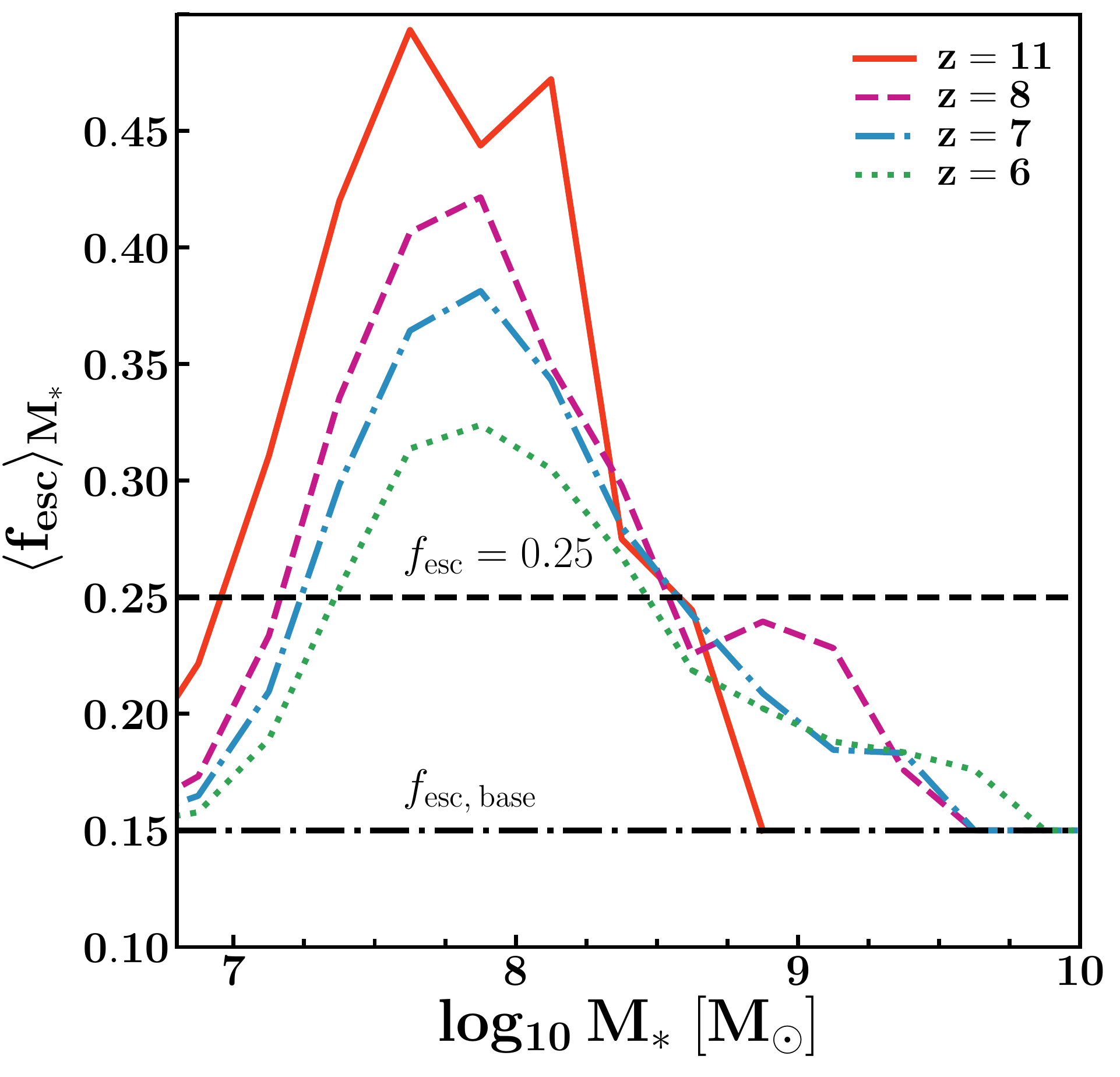}
	\caption{Mean escape fraction within each stellar mass bin for the quasar activity scenario.  As redshift decreases the dynamical time of the halo, which controls how long the boosted value of $f_\mathrm{esc}$ lasts for, increases causing the distribution to widen.  Alongside this increase in dynamical time, hierarchical assembly causes the stellar mass budget to be dominated by higher mass galaxies leading to an associated evolution in the ionizing emissivity, as shown in Figure \ref{fig:Nion}.}
	\label{fig:fescMvir}
\end{figure}

Using our toy model in which the value of $f_\mathrm{esc}$ is boosted following a quasar wind event ejecting all galaxy gas, in Figure \ref{fig:fescMvir} we show the mean value of $f_\mathrm{esc}$ as a function of galaxy stellar mass, $\langle f_\mathrm{esc}\rangle_{\mathrm{M}_*}$, at various redshifts.  Perhaps the most striking feature is that $\langle f_\mathrm{esc}\rangle_{\mathrm{M}_*}$ peaks for galaxies with stellar mass $\mathrm{M}_\mathrm{*} = 10^{7.5}$-$10^{8.0} \mathrm{M}_\odot$.  This is a result of merger rates, gas fractions and black hole masses conspiring to cause quasar events to be the most numerous at this mass scale.  At lower stellar masses, galaxies tend to be satellites and hence do not have quasar events (quasars are only triggered in the central galaxy of a merger event). While for $\mathrm{M}_* \gtrsim 10^8 \mathrm{M}_\odot$, quasar events do not have enough energy to eject the copious gas reservoirs, reducing the likelihood of $f_\mathrm{esc}$ being boosted for galaxies above this mass scale.

The second noticeable feature in Figure \ref{fig:fescMvir} is that the $\langle f_\mathrm{esc}\rangle_{\mathrm{M}_*}$ distribution begins to widen and flatten as redshift decreases.  This is a consequence of the dynamical time of dark matter halos increasing with decreasing redshift, causing ejected gas to take longer to be reincorporated back into the galaxy.  Hence, as redshift decreases, the duration of free photon escape following a quasar ejection event increases.  In conjunction with hierarchical assembly creating more massive objects, this causes the distribution to widen and flatten.  Eventually the stellar mass budget is dominated by objects beyond the peak of the $\langle f_\mathrm{esc}\rangle_{\mathrm{M}_*}$ distribution. However, it is not until redshift $z \leq 8$ that the mean $f_\mathrm{esc}$ falls below $0.20$, causing the two ionizing emissivities to be almost identical until they begin to diverge slightly at very late times (see also	 Figure \ref{fig:Nion} below $z \approx 8$).

This bias of the $\langle f_\mathrm{esc}\rangle_{\mathrm{M}_*}$ distribution towards intermediate mass galaxies could have important consequences on the $21$cm power spectrum.  During the Epoch of Reionization, the size of ionized hydrogen regions depends upon the number of ionizing photons that escape into the IGM.  In the constant $f_\mathrm{esc}$ scenario, the number of ionizing photons is directly proportional to the galaxy stellar mass.  Therefore, at a fixed neutral hydrogen fraction, we find a large number of small ionized regions surrounding low mass galaxies in tandem with a handful of large ionized regions around the higher mass objects.  From Figure \ref{fig:fescMvir}, we see that boosting $f_\mathrm{esc}$ following quasar activity is most efficient in intermediate mass objects.  Hence, under the quasar activity scenario, we would expect a relative increase (decrease) in the number of large (small) ionized regions.  Such a difference would have a direct impact on the $21$cm power spectrum with the power at being diminished at large scales and enhanced at small scales compared to a constant $f_\mathrm{esc}$ scenario.  This presents a wealth of opportunities to explore how the $21$cm power spectrum responds to physically motivated $f_\mathrm{esc}$ scenarios.  For example, supernova feedback is thought to regulate the value of $f_\mathrm{esc}$ in low mass galaxies \citep{Trebitsch2017,Kimm2017}, which in combination with our quasar activity scenario would result in an increased value of $f_\mathrm{esc}$ for both intermediate and low mass galaxies.  Investigating the impact of different physically motivated $f_\mathrm{esc}$ scenarios on the reionization topology will greatly aid ongoing and future $21$cm signals and high-redshift galaxy observations. However, such an analysis requires a suite of reionization simulations, ideally with the effect of reionization self-consistently coupled to galaxy evolution which is beyond the scope of this letter. We plan to address this analysis in future work.

\section{Summary} \label{sec:Conclusion}

In this letter we have presented an idealized reionization model in which the dispersion of gas clouds following quasar wind ejection allows ionizing photons from stars to escape into the IGM more easily.  Whilst the ionizing emissivity evolves differently for this scenario compared to a constant value of $f_\mathrm{esc}$, the difference is mostly noticeable above redshift $z \geq 10$.  Below this redshift, the dynamical time of the host dark matter halo is large enough to delay reincorporation of ejected gas and extend the period of easy photon escape.  We find that under this quasar activity scenario, the mean value of $f_\mathrm{esc}$ remains close to the baseline value for low stellar mass galaxies, peaks for intermediate mass and has a gradual decline for higher mass galaxies,  only again reaching the baseline value for the rarest objects.  This peak for intermediate mass galaxies could have a significant effect on the $21$cm power spectrum through the enhancement and reduction of power on small and large scales.  Our model shows that even under the simplest of assumptions, the influence of quasars on reionization could go beyond their intrinsic emission of ionizing photons.  It is an exciting prospect that radio telescopes such as the Square Kilometre Array and the Hydrogen Epoch of Reionization Array could provide $21$cm power spectrum measurements and potentially constrain the impact of quasars on the reionization topology.

\section*{Acknowledgements}

We would like to thank Emma Ryan-Weber for useful discussion and comments.  JS and AH are supported under the Australian Research Council's Discovery Project funding scheme (project number DP150102987). Parts of this research were conducted by the Australian Research Council Centre of Excellence for All Sky Astrophysics in 3 Dimensions (ASTRO 3D), through project number CE170100013.  The Semi-Analytic Galaxy Evolution (\texttt{SAGE}) model used in this work is a publicly available codebase that runs on the dark matter halo trees of a cosmological N-body simulation.  It is available for download at \url{https://github.com/darrencroton/sage}.

\bibliographystyle{mnras}
{\footnotesize
	\setlength{\itemsep}{1pt}
	\begin{spacing}{0.01}
		\bibliography{references}
	\end{spacing}	
}

\end{document}